\documentclass[twocolumn,prl,superscriptaddress,showpacs,preprintnumbers,amsmath,amssymb]{revtex4}

\usepackage{graphicx}
\usepackage{dcolumn}
\usepackage{bm}

\begin{document}


\title{Frustration-Driven Successive Metamagnetic Transitions in TbB$_4$}

\author{T. Inami} \author{K. Ohwada}
\affiliation{%
Synchrotron Radiation Research Unit, Japan Atomic Energy Agency, Sayo, Hyogo
679-5148, Japan
}%

\author{Y. H. Matsuda} 
\author{Z. W. Ouyang} 
\author{H. Nojiri}%
\affiliation{
Institute for Materials Research, Tohoku University, 
2-1-1 Katahira, Sendai 980-8577, Japan
}%

\author{D. Okuyama}
\author{T. Matsumura}
\author{Y. Murakami}
\affiliation{
Department of Physics, Faculty of Science, Tohoku University, 
Aoba, Sendai 980-8578, Japan
}%

\date{\today}%

\begin{abstract}
  Resonant magnetic x-ray diffraction experiments on the
  Shastry-Sutherland lattice TbB$_4$ were carried out under strong
  pulsed magnetic fields up to 30~T. TbB$_4$ exhibits a multi-step
  magnetization process above 16~T when magnetic fields are applied
  along the $c$-axis. We examined the intensity of the 010 magnetic
  reflection as a function of magnetic field and found that the
  magnetization plateau phases are accompanied by large XY components
  of magnetic moments, in contrast to normal fractional magnetization
  plateau phases. The magnetization was calculated using a simple spin
  model deduced from the above result. Finally we propose that
  frustration is the key to understanding the observed multi-step
  magnetization.
\end{abstract}

                              
\maketitle

When one applies pressure to a substance, the volume will be reduced
at all times depending on its compressibility. In contrast to such
natural behavior, several magnetic materials exhibit magnetization
plateaus, where the magnetization keeps a constant value at a finite
range of magnetic fields even below the saturation field. This ``zero
compressibility'' is simply ascribed to the fact that $S_{\it z}$ is a
good quantum number, while a variety of mechanisms give rise to
plateaus in the magnetization curve.


Let us consider only plateau phases in ordered states here. The
easy-axis magnetization of an Ising antiferromagnet is most
comprehensible. An example is the one-third magnetization plateau of
CoCl$_2$$\cdot$2H$_2$O~\cite{cocl2}. Competing exchange interactions
stabilize the $\uparrow\uparrow\downarrow$ structure just below the
saturation field. The multi-step magnetization curves of
PrCo$_2$Si$_2$~\cite{prco2si2} and CeSb~\cite{cesb} are also
classified into this category. The plateau phases are described by a
complex sequence of up- and down-spins.

Magnetization plateaus are also observed in Heisenberg spins. In
geometrically frustrated Heisenberg antiferromagnets, one-third
($\uparrow\uparrow\downarrow$) and one-half
($\uparrow\uparrow\uparrow\downarrow$) plateaus are sometimes observed
in triangular- and tetrahedon-based lattices, respectively. Although
the field range of the plateau phase is zero for these lattices if
only the nearest-neighbor interaction is taken into account,
additional mechanisms, such as a four-spin exchange coupling, quantum
fluctuations, single-ion anisotropy and magnetoelastic couplings,
stabilize the plateau phase in real
magnets~\cite{c6eu,rfmo,gdpd2al3,cdcr2o4}. A crucial point is that all
plateau phases have a collinear magnetic structure. No ordered
magnetic moment perpendicular to the magnetic field exists in the
plateau phases.

Recently, a multi-step magnetization curve was observed in TbB$_4$.
Intriguingly, several experimental evidences indicate that TbB$_4$
resembles none of the above-mentioned magnets. In this letter, we
report on resonant x-ray diffraction (RXD) experiments on TbB$_4$ up
to 30~T. The field dependence of the magnetic intensity provided
decisive information on the magnetic structure of the field-induced
phases. From this result, we show that TbB$_4$ indeed belongs to
a rare class of magnets, characterized by ordered perpendicular
components in the plateau phases. We also show that frustration plays
a vital role in the multi-step magnetization process on the basis of
the calculation of magnetization curves.

The rare-earth tetra-boride $R$B$_4$ crystallizes into a tetragonal
structure (space group: $P4/mbm$). The network of the rare-earth ions
is equivalent to the frustrated Shastry-Sutherland (SS)
lattice~\cite{SS}. Several pieces of work have thus been carried out
from a viewpoint of frustration between rare-earth quadrupole
moments~\cite{wata,matsux}. Somewhat confusingly, in the strict sense,
classical continuous spins on the SS lattice are not frustrated,
because the ground state is a unique helical or N\'eel structure and
has no degeneracy~\cite{SS}. However, for helical structures, none of the
antiferromagnetic bonds are antiparallel and thus local pair
exchange-energy is not minimized. The spins are not satisfied and still
feel ``frustration''. We will revisit this point later.

\begin{figure}
\scalebox{0.45}{\includegraphics{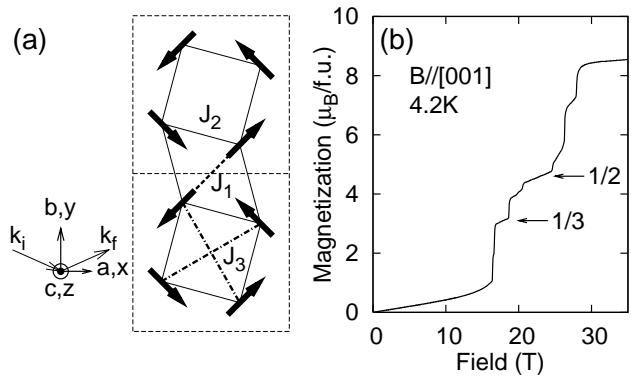}}
\caption{\label{fig:2}
  (a) Magnetic structure of TbB$_4$ between $T_{\rm N1}$ and $T_{\rm
    N2}$. A large square is the unit cell. Major interactions ($J_1$ :
  broken line, $J_2$ : solid line, $J_3$ : dot-dashed line) are shown.
  Sample coordination ($a,b,c,x,y$ and $z$) and incident and scattered
  x-rays ($k_{\rm i}$ and $k_{\rm f}$) are also shown. (b)
  Magnetization process of TbB$_4$ taken from Ref.~\cite{yoshii}.}
\end{figure}

In TbB$_4$, there are two transition points $T_{\rm N1}$=44~K and
$T_{\rm N2}$=22~K. The total angular momentum of the Tb ion is large
($J$=6) and hence the Tb spin is approximated by a classical spin. The
magnetic structure between $T_{\rm N1}$ and $T_{\rm N2}$ revealed by
neutron powder diffraction experiments is shown in
Fig.~\ref{fig:2}(a)~\cite{matsun}. The magnetic moments are XY-type
and are confined in the basal $ab$-plane. Additional in-plane
anisotropy orients the moments parallel to the diagonal line. The
nearest-neighbor interaction $J_1$ is antiferromagnetic. The
next-nearest-neighbor interaction $J_2$ is probably antiferromagnetic,
and $J_1$ and $J_2$ form a frustrated SS lattice. In $R$B$_4$, the
further-neighbor antiferromagnetic interaction $J_3$ is taken into
account, which stabilizes the observed magnetic structure. In this
structure, the $J_2$ bond is not satisfied. Below $T_{\rm N2}$, the
magnetic moments rotate in the $ab$-plane and a collinear-like
antiferromagnetic structure is realized. This transition is considered
to be a ferro-quadrupole order~\cite{fujita} and is destroyed by
application of magnetic fields.

The magnetization process of TbB$_4$ parallel to the $c$-axis shown in
Fig.~\ref{fig:2}(b) exhibits multi-step metamagnetic behavior between
16 and 28~T~\cite{yoshii}. The number of the steps is reported to be
nine, and the one-half magnetization plateau is most prominent. When a
magnetic field is applied parallel to the [100] and [110] directions, a
single metamagentic transition is observed. Magnetization measurements
of diluted samples indicated that the multi-step magnetization is
inherent to the phase between $T_{\rm N1}$ and $T_{\rm
  N2}$~\cite{iga}. The metamagnetic behavior suggests that the
essential nature of the transitions is level-crossing between
low-lying multiplets~\cite{tb}. However, the origin of the multiple
successive transitions is not easily understood. Magnetic structures
in the high-field plateau phases provide useful information, thus we
conducted an x-ray diffraction experiment under strong pulsed magnetic
fields, which has been recently developed utilizing brilliant
third-generation synchrotron x-rays~\cite{matsuda2,narumi,esrf}.
Although magnetic x-ray scattering cross-sections are very small, we
utilized resonant enhancement of magnetic scattering~\cite{Ho}.

Synchrotron x-ray diffraction experiments were carried out at
undulator beamline BL22XU at SPring-8. The x-ray energy was tuned to
the Tb $L_{\rm III}$ absorption edge (7.514~keV). Magnetic fields up
to 30~T were generated using a small pulsed magnet (20~mm in outer
diameter and 24~mm in length). The pulse duration was about 0.6~ms.
The magnet was a split-pair, and two windows were prepared for
incoming and outgoing x-rays. The magnet and a sample were attached to
the 100~K and 10~K stages of a conventional closed cycle refrigerator,
respectively. Details of experimental setup are described in
Ref.~\cite{matsuda2}. Single crystals of TbB$_4$ were grown by the
floating zone method. The sample was a thick plate of dimensions
1~mm$\times$1~mm$\times$0.5~mm with a polished (010) surface. The FWHM
of the $\omega$ profile of the 020 reflection was about 0.06$^\circ$.

The scattering plane was horizontal. The $b^*$ axis lay in the
scattering plane and the $c$-axis was perpendicular to the scattering
plane. We observed the forbidden 010 reflection, which is a magnetic
Bragg point as demonstrated in the neutron experiment~\cite{matsun}.
The azimuthal angle $\Psi$, which is the rotation about the scattering
vector, is a key parameter in RXD. We define $\Psi$=0 when the
$c$-axis is perpendicular to the scattering plane. $\Psi$ was kept
close to zero throughout this experiment. Magnetic fields were applied
parallel to the $c$-axis. We used a high-counting-rate detector and
measured diffraction intensity as a function of time with a typical
resolution of 10~$\mu$s. The field strength was also monitored as a
function of time, and we obtained the field dependence of the
diffraction intensity from these two series of data.  Polarization of
scattered x-rays were separated using a PG (006) crystal analyzer. In
order to prevent the sample from being heated by x-ray irradiation, we
inserted a chopper into incident x-rays. The window width was about
2~ms and the duty cycle was about 0.01.

\begin{figure}
\scalebox{0.5}{\includegraphics{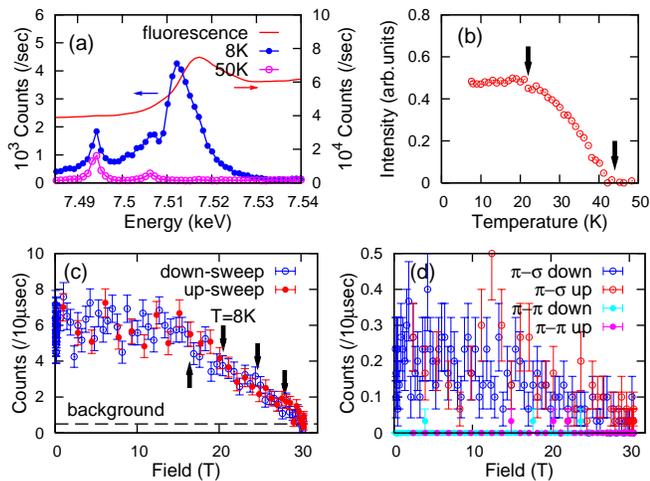}}
\caption{\label{fig:3} 
  (color online) (a) Peak intensity of the 010 reflection as a
  function of photon energy below $T_{\rm N2}$ and above $T_{\rm N1}$.
  Huge enhancement is evident at the $L_3$ edge of the fluorescence
  spectrum. Small peaks observed at 50~K are multiple scattering. (b)
  Intensity of the 010 reflection as a function of temperature.
  Arrows indicate $T_{\rm N1}$=44~K and $T_{\rm N2}$=22~K. The sample
  temperature is higher than the thermometer by 2~K because of beam
  heating. (c) Field dependence of the peak intensity of the 010
  reflection at 8~K. The arrows indicate the onset of the high-field
  phases (16.4~T), the lower and upper critical fields of the one-half
  plateau phase (20.5~T and 24.7~T) and the saturation field (28~T) at
  4~K. (d) Field dependence of the 010 reflection at 8$\sim$10~K.
  Polarization of scattered x-rays are separated. Red and blue
  (magenta and cyan) circles are up- and down-sweeps of the
  $\pi$-$\sigma$ ($\pi$-$\pi$) channel.}
\end{figure}

In Fig.~\ref{fig:3}(a), we show the peak intensity of the 010
reflection as a function of photon energy below $T_{\rm N2}$ and above
$T_{\rm N1}$. The fluorescence spectrum is also shown. The intensity
of the 010 reflection exhibits huge enhancement at the $L_{\rm 3}$
main edge, implying that the resonance is ascribed to electric dipole
(E1) transition. Subsequent measurements were carried out at the peak
energy (7.512~keV). The temperature dependence of the integrated
intensity of the 010 reflection is shown in Fig.~\ref{fig:3}(b). The
intensity appears below $T_{\rm N1}$ and is constant below $T_{\rm
  N2}$. The effect of beam heating is estimated to be at most 2~K from
the differences between the reported and observed transition
temperatures.

Here we briefly mention the resonant x-ray scattering
amplitude~\cite{blume} of the Tb ion, which is proportional to
\begin{equation}
  \label{eq:1}
f_{\rm res} \propto i C_1 ({\bm\epsilon}_{\rm f}^*\times
{\bm\epsilon}_{\rm i})\cdot{\bm m}+ 
C_2\,{\bm\epsilon}^\dag_{\rm f}\,O\,{\bm\epsilon}_{\rm i},
\end{equation}
where $\bm m$ is the magnetic moment and ${\bm\epsilon}_{\rm i}$ and
${\bm\epsilon}_{\rm f}$ are the polarization vectors of the incident
and scattered x-rays, respectively. The symmetric second-rank tensor
$O$ describes anisotropy of the Tb 5$d$ orbital caused by anisotropic
crystal environment or quadrupole order and is represented by a linear
combination of the five elements $O_{xy}$, $O_{yz}$, $O_{zx}$,
$O_{22}$ and $O_{20}$. The first term in eq.(\ref{eq:1}) is magnetic
scattering and the last term causes anisotropy of the tensor of
susceptibility (ATS) scattering. Thorough measurements on temperature
and azimuthal angle dependence with polarization analysis were
performed by TM~\cite{matsuxx}. The results were consistent with the
magnetic structure obtained in the neutron experiment~\cite{matsun},
and it was confirmed that 0$k$0 reflections ($k$ : odd integer) at
$\Psi$=0 are magnetic scattering.

The field dependence of the peak intensity of the 010 reflection at
8~K is shown in Fig.~\ref{fig:3}(c). The data were averaged over 12
field-scans. The intensity is constant up to 16~T and gradually
decreases above 16~T as the field is increased. An important
outcome is that a considerable amount of the intensity remains in the
magnetization plateau phases above 16~T. From this result, we
understand that the plateau phases include (0,1,0) modulated magnetic
moments and/or anisotropic 5$d$ orbitals.

The second important outcome is obtained from polarization dependence
of scattered x-rays. States of linear polarization are labeled by
$\sigma$ and $\pi$ when they are normal and parallel to the scattering
plane, respectively. The incident polarization was $\pi$. The
scattering amplitudes $f_{\rm res}(\theta)$ of $\pi\pi^\prime$ and
$\pi\sigma^\prime$ channels for each magnetic and ATS component at
$\Psi$=0 are listed in Table \ref{tab:1}, where $\theta$ is the Bragg
angle. The structure factor $f_{\rm res}(\theta)\sum_{\rm
  i}e^{i2\pi\bm{KR}_{\rm i}}$ at (0,$k$,0) is proportional to either
$if_{\rm res}(\theta)\sin 2\pi k \delta$ or $f_{\rm res}(\theta)\cos
2\pi k \delta$, where the fractional coordinate of the Tb site
$\delta\sim0.317$. The structure factor below 16~T is proportional to
$m_x \cos\theta \sin 2\pi k \delta$. In Fig.~\ref{fig:3}(d), field
dependence of $\pi\pi^\prime$ and $\pi\sigma^\prime$ channels of the
010 reflection are shown. The data were averaged over 30 field-scans.
It is obvious that the scattered x-rays are dominated by
$\pi\sigma^\prime$ channel at all field ranges. Hence it is inferred
from Table \ref{tab:1} that the scattering intensity of the 010
reflection arises from $m_x$+$m_y$ or $O_{yz}$+$O_{zx}$. We also
observed the 050 reflection (not shown) without the analyzer. In spite
of large difference in $\theta$ (6.6$^\circ$ and 35.2$^\circ$), the
field dependence of the 050 reflection is very similar to that of the
010 reflection. This fact indicates that the intensities of both the
zero- and high-field phases are proportional to $\cos\theta\sin2\pi k
\delta$ and that $m_z$ ($\propto\sin2\theta$) does not exist in the
high-field phases. In addition, recent neutron diffraction experiments
of TbB$_4$ under pulsed magnetic fields have shown that the field
dependence of the 010 reflection in neutron diffraction is also
similar to that of x-ray diffraction~\cite{yoshiin}. Neutrons observe
$m_x$, $m_y$ and $m_z$. Hence pure $O_{yz}$+$O_{zx}$ order is
inconsistent with the neutron result. $m_x$ or $m_y$ must exists.
Actually all data are well interpreted by assuming that the major
order parameter in the high-field phases is $m_x$, magnetic moments
perpendicular to magnetic fields.

\begin{table}[t]
  \centering
  \begin{tabular}[t]{ccccccccc} \hline
& $m_x$ & $m_y$ & $m_z$ & 
$O_{xy}$ & $O_{yz}$ & $O_{zx}$ & $O_{22}$ & $O_{20}$ \\ \hline
${\pi\pi^\prime}$     & 0 & 0 & $i\sin2\theta$ & 
                     0 & 0 & 0 & $-1$ & $-\cos2\theta$ \\
${\pi\sigma^\prime}$ & $i\cos\theta$ & $-i\sin\theta$ & 0 & 
                     0 & $\cos\theta$ & $\sin\theta$ & 0 & 0 \\ \hline
  \end{tabular}
  \caption{Scattering amplitude of each component of magnetic and 
    ATS scattering at $\Psi$=0 for $\pi\pi^\prime$ and $\pi\sigma^\prime$ 
    channels.}
  \label{tab:1}
\end{table}
 
This is a rather surprising result, because all magnetic moments
in plateau phases are parallel or antiparallel to the magnetic field
as mentioned in the introduction. A canted magnetic structure, usually
derived from the coexistence of parallel and perpendicular components,
results in a continuous magnetization curve unlike the observed
step-like magnetization process. Accordingly, we submit an idea that
in TbB$_4$ the magnetic structures in the plateau phases consist of
very hard XY-type magnetic moments and Ising-like magnetic moments.

In order to interpret this novel magnetic structure, we propose the
following spin model. The low-lying multiplets are approximated by the
ground-state XY-spin $\vec S$ with the in-plane anisotropy $E$ and the
excited-state Ising spin $\vec s$ with the energy gap $G$. The spin
Hamiltonian is
 \begin{eqnarray*}
 H&=&\sum_{m=1,2,3}J_m
 \left(\sum_{i,j}\left(S_i^x S_j^x + S_i^y S_j^y\right) + 
 \sum_{k,l}s_k^z s_l^z\right)\\
 &-& E\sum_i(|S_i^x S_i^y|-\frac{1}{2}) + G\sum_k(s_k^z)^2 -
 g\mu_{\rm B}H_z\sum_k s_k^z,
 \end{eqnarray*}
where the spins are unit vectors, $i$ and $j$ sites are occupied by
XY spins and $k$ and $l$ sites are occupied by Ising spins.

\begin{figure}
\centering
\scalebox{0.5}{\includegraphics{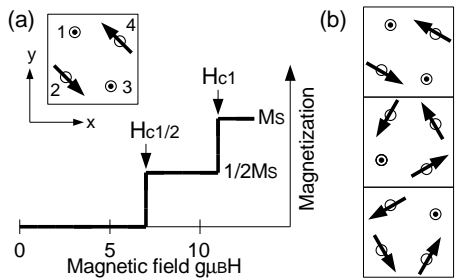}}
\caption{\label{fig:4} 
  (a) Calculated magnetization curve for $E\rightarrow\infty$. The
  magnetic structure of the one-half plateau phase is shown in the
  inset.  $\odot$ denotes a spin parallel to the $z$-axis. (b)
  Possible magnetic structure of the one-third plateau phases.}
\end{figure}

Magnetization curves were calculated for finite-size cells with
periodic boundary conditions. We consider 4 spins in the
crystallographic unit cell ($1\times 1$ cell). For simplicity, we
initially consider a four-clock model by taking the limit
$E\rightarrow\infty$. For example (see Fig.~\ref{fig:4}(a)), allowed
states for spins 1 and 3 are $\swarrow
(-\frac{\sqrt{2}}{2},-\frac{\sqrt{2}}{2},0)$, $\nearrow
(\frac{\sqrt{2}}{2},\frac{\sqrt{2}}{2},0)$, $\odot$(0,0,1) and
$\otimes$(0,0,$-1$). In this case, we calculated energies of all
$4^4$=256 states and found ground states. For rather wide parameter
sets, this model has the same ground state as observed. A
calculated magnetization curve for $J_1=2, J_2=1, J_3=2$ and $G=1$ is
shown in Fig.~\ref{fig:4}(a). The one-half magnetization plateau is
clearly reproduced and the corresponding magnetic structure is shown
in the inset. Similar results were obtained with other sets of
parameters and different cell sizes ($2\times2$ and $1\times3$ cells).

It is interesting to note that a number of states that have various
magnetizations are degenerate at the critical fields $H_{\rm c1/2}$
and $H_{\rm c1}$. This is in fact an ordinary phenomenon observed in
metamagnetic transitions. However, if this degeneracy is lifted, the
multi-step magnetization curve of TbB$_4$ can be reproduced. Three
mechanisms can lift the degeneracy: (i) difference in
magnitude between $S$ and $s$, (ii) interactions beyond $J_3$, and
(iii) frustration. While the first two mechanisms cause symmetric
splits at $H_{\rm c1/2}$ and $H_{\rm c1}$, the observed magnetization
curve is asymmetric. For instance, the one-third plateau is clearly
visible, whereas the two-third plateau is indistinguishable, as seen
in Fig.\ref{fig:2}(b). It is evident that the last mechanism,
frustration, plays an essential role. A possible magnetic structure of
the one-third plateau phase is shown in Fig.~\ref{fig:4}(b). Here we
set the in-plane anisotropy $E$ finite so that the moments can rotate
in the $ab$-plane. The Ising sites behave like defects. In frustrated
magnets, as pointed out in the introduction, the local pair
exchange-energy is not minimized. Hence, when a defect is introduced,
the magnetic structure around the defect relaxes and gains exchange
energy. This does not happen in magnetic lattices without frustration,
because all exchange energies have been minimized.

For magnetic systems with short-range interactions, it is expected
that small commensurate unit cells are more stable than large or
incommensurate cells. Thus it is plausible that a finite number of
selected magnetization steps appear in the magnetization curve through
the frustration mechanism. This mechanism is active only when the
density of the Ising sites is less than half (up to $J_3$). Therefore
the degeneracy is lifted only at $H_{\rm c1/2}$. The diffraction
intensity of magnetic structures based on the model shown in
Fig.~\ref{fig:4}(a) is one-quarter of the intensity at 0~T. The
observed intensity is, however, about 40\% of that at 0~T, implying
presence of other order parameters, such as $O_{20}$ and $O_{yz}$.


In conclusion, we have investigated the multi-step magnetization
process of TbB$_4$ using RXD. The magnetic intensity observed in the
plateau phases illustrates that large XY-components of the magnetic
moments coexist with aligned Z-components. A simple spin model
reproduces the one-half magnetization plateau, and it is found that
the frustrated nature of the SS lattice lifts the degeneracy at the
critical fields and creates the observed multi-step metamagnetic
behavior. Paradoxically, frustration is not only an origin of
degenerate states, but also breaks degeneracy.

This work was supported by Grant-in-Aid for Scientific Research on
Priority Areas ``High Field Spin Science in 100T'' (No.451) from the
Ministry of Education, Culture, Sports, Science and Technology (MEXT).


\end{document}